# Augmented reality applications in manufacturing and its future scope in Industry 4.0

Omid Ziaee, Mohsen Hamedi

**Abstract**

Augmented reality technology is one of the leading technologies in the context of Industry 4.0. The promising potential application of augmented reality in industrial production systems has received much attention, which led to the concept of industrial augmented reality. On the one hand, this technology provides a suitable platform that facilitates the registration of information and access to them to help make decisions and allows concurrent training for the user while executing the production processes. This leads to increased work speed and accuracy of the user as a process operator and consequently offers economic benefits to the companies. Moreover, recent advances in the internet of things, smart sensors, and advanced algorithms have increased the possibility of widespread and more effective use of augmented reality. Currently, many research pieces are being done to expand the application of augmented reality and increase its effectiveness in industrial production processes. This research demonstrates the influence of augmented reality in Industry 4.0 while critically reviewing the industrial augmented reality history. Afterward, the paper discusses the critical role of industrial augmented reality by analysing some use cases and their prospects. With a systematic analysis, this paper discusses the main future directions for industrial augmented reality applications in industry 4.0. The article investigates various areas of application for this technology and its impact on improving production conditions. Finally, the challenges that this technology faces and its research opportunities are discussed.

**Introduction**

Nowadays, the manufacturing industry is undergoing an industrial revolution deemed industry 4.0 (I4.0) (Ślusarczyk, Haseeb and Hussain, 2019). As illustrated in Figure 1, the first industrial revolution occurred with the mechanization of processes, followed by mass production in industry 2.0 and automation in industry 3.0 (Ababsa, 2020). Industry 3.0 introduced the swarm of robots and sensors to the shop floor and automated many manufacturing activities. I4.0 is known as the new industrial stage where the integration of Industrial Production Systems (IPS) and Information and Communication Technologies (ICT), especially Internet of Things (IoT), happens. This integration forms a Cyber-Physical Space (CPS) (Wang et al., 2015; Jeschke et al., 2016, pp. 3–19) and improves manufacturing operations and products (Dalenogare et al., 2018). The I4.0 paradigm will reshape the way operators do the tasks by making them entirely connected and highly automated through CPS integration of the tools, devices, and machines.

Augmented Reality (AR) is known as one of the main pillars of the I4.0 paradigm (Erboz, 2017), and it can bridge the gap between the real and the progressively crucial digital environment for its user (Masood & Egger, 2019). Furthermore, I4.0 is founded upon technologies with two aspirations: 1. Accelerating the remaining section of manufacturing activities that needs human expertise and skills and 2. Integrating intelligence into the automated infrastructure makes autonomous robotic systems able to make decisions from data (Lorenz et al., 2105). The first aspiration made the provisions for incorporating AR into the manufacturing field. The second aspiration built the foundations for incorporating AI into manufacturing (Sahu et al., 2020). It is hard to find a commonly accepted definition for AR due to its wide use in various fields (Liao, 2016). Yet, early academic research defined

AR as a system that combines the real and virtual environment, registered in three dimensions, and allows their interactivity in real-time (Azuma, 1997).

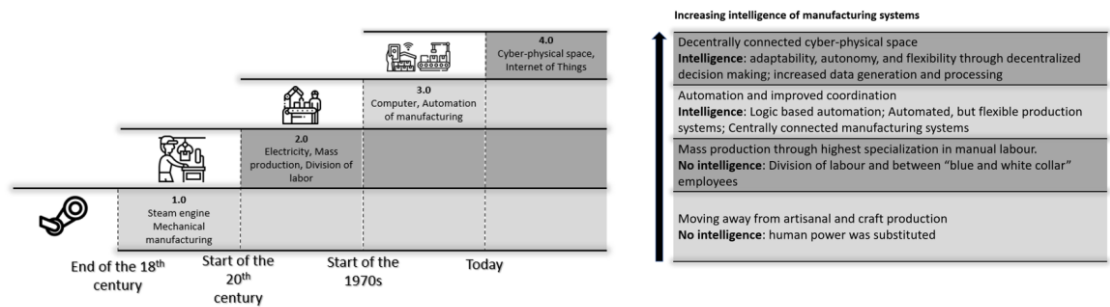

Figure 1. The four industrial revolutions

Source: reproduced from (Egger & Masood, 2019), images: Flaticon.com

## 1. Literature review
### 1.1. Industrial Augmented Reality

In industry, the need for the mean that connects the cyber and physical space led to the development of the Industrial Augmented Reality (IAR) technologies, where AR acts as an interface for human-machine communications (Ong & Nee, 2004). To present a brief history of IAR, we discuss its milestones related to production systems. This technology was first used in 1965 when Ivan Sutherland developed the first head-mounted display system, which illustrated the user a simple wireframe drawing rotating in space according to their angle of view in the user field of view (FOV). In 1974, Myron Krueger created a device called Videoplace, which can be considered the first simultaneous use of AR technology and the human-machine interface (HMI). In this system, the user's hands' position was closely monitored by the camera, and the person was able to work according to the options on the screen wirelessly without using any controller. Sixteen years later, in 1990, Caudell and Mizell at Boeing first coined the term augmented reality. They used an AR system that could assist wiring operators by dynamically displaying potential drill holes in an airplane fuselage on a Head-Mounted Device (HMD), which guided operators to perform their operations in less time and more accurately without using a handbook or template (Caudell & Mizell, 1992). In 1992, Lewis Rosenberg designed a US Air Force device that could be used to train personnel to work with remotely controlled robots (Rosenberg, 1992). In 1998, Sportvision first used AR for entertainment. The aim was to make it easier to watch the American football game; the company overlaid a yellow line on the first and tenth lines in the cameras' images. This system was based on laser scanning of the playing field and taking precise location information of the cameras, and getting the scoring line's location by its operator (Lake, 2000). In 1999, Kato introduced an open-source software for processing received images by detecting markers and HMD calibration for image placement named ARToolkit (Kato & Billinghurst, 1999). This algorithm is still used after about 20 years in cases where the AR system works based on detecting a marker or barcode (Khan et al., 2018; Blanco-Pons et al., 2019). From 1999-2003, the ARVIKA project made IAR gain significant attraction. ARVIKA was a research project funded by BMBF (Germany Federal Ministry of Education and Research), where a consortium consisting of 23 partners from industry, research institutions, and universities worked on AR development for industrial applications (Weidenhausen et al., 2003). ARVIKA developed the AR applicability, and its positive result was evident in later projects like ARTESAS (Advanced Augmented Reality Technologies for Industrial Service Applications) in 2004-2006. ARTESAS, in 2007 further developed the IAR for use in automobiles, aircraft, and the automation industry (Wohlgemuth & Friedrich, 2007).

In 2013, Volkswagen launched the MARTA app (Mobile Augmented Reality Technical Assistance), which primarily gave technicians step-by-step instructions for repair operations. 2014 was a turning point in the history of AR, as Google was able to attract a broad audience's attention by introducing their AR glasses to society. However, the project has been shut down mainly due to marketing reasons

(Haque, 2015; Reynolds, 2015). In 2015 and 2016, major companies such as Microsoft and Magic Leap launched their new HMD, which can be considered pioneers in using AR in the industry. In these products, their weight decreased, and their internet connection, as well as the image quality, improved. With the introduction of Microsoft HoloLens 2 in 2019, this technology in the industry expanded. HoloLens2 can detect the user's hands and enable them to interact with virtual objects, and support several ready-made programs for various applications. With the emergence of more powerful HMDs and AR Enterprise Platforms like Atheer, Scope AR, Ubimax, and Upskill, both the hardware and software portions of this technology rapidly developed for industrial purposes.

### 1.2. IAR current achievements and future prospects

Currently, many enterprises are using IAR in various fields. Boeing has developed Tom Caudell's work, who claims that with the help of this technology, he has increased aircraft wiring operations speed by 25% and reduce the number of errors by 50% (Gigazine, 2016). NGRAIN reports that in partnership with Lockheed Martin, it has used AR technology to increase staff speeds by up to 30% and accuracy by up to 96% in the F-35 production process (Alexander George, 2015). With the help of Microsoft HoloLens glasses, ThyssenKrupp has succeeded in improving the ordering and production process of its product's staircase devices and bringing orders to 4 times earlier (Bardeen, 2017). With AR's help, Airbus personnel put the brackets installed on the plane on the digital image, compare it with the predetermined model, and inform the operator of the existing problems. This is done with a tablet that uses image processing to identify each bracket and check its appearance. Due to the high sensitivity of the inspection work, it used to take three weeks to check an aircraft's brackets before using this AR technology, but now this time has been reduced to three days (Airbus, 2017). DHL's report on the impact of using AR in the warehousing sector states that using AR technology enabled them to reduce staff error by up to 40% (DHL, 2014). Another report of a pilot of DHL and RICOH claims that using AR, order picking speed increased 25%, while the feedback from the users was quite favorable (Kückelhaus, 2016). In the case of General Electric Health Company, the productivity in the warehousing operations also has increased by 46% (Abraham & Annunziata, 2017). Also, in the GE Aviation report, increased efficiency of 8-12% is observed, saving millions of dollars for engine maintenance tasks. The maintenance workers get instructions on the optimized B-nut tightening procedure from an HMD that receives the torque inserted into the B-nut from a Wi-Fi-enabled wrench. The worker's compliance and job satisfaction are also improved (Martin, 2017). (Tang et al., 2003) reports thattheir AR-assisted assembly operations have registered a reduction of error of up to 82% for their experimental research.

AR's total addressable market is forecast to expand during the COVID-19 pandemic, growing from $15.3 billion in 2020 to reach $77 billion in 2025 (MarketsandMarkets, 2020), registering a five-year Compound annual growth rate (CAGR) of 38.1%. Goldman Sachs Research expects this number to exceed $80 billion in the same year (Goldman Sachs Research, 2016). This promising trend is also seen for the IAR in the automotive industry, where its global AR market size is projected to reach USD 2651.6 million by 2026, from USD 2393.4 million in 2020, with a CAGR of 10.0% (QYResearch, 2020). AR significantly impacts the manufacturing industry (JABIL, 2018). Many researchers claim that more work is needed to implement IAR in industries entirely (Scurati et al., 2018; Gattullo et al., 2019; Mourtzis et al., 2019; Nee et al., 2012).

### 1.3. AR and its working principle

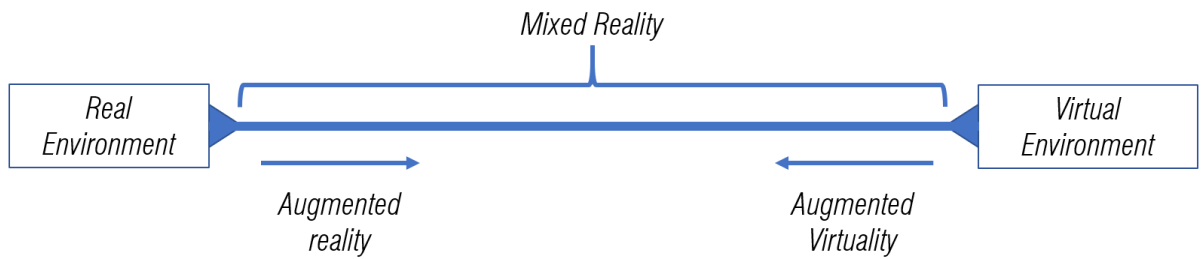

Figure 2. Modified RV

Source: Based on (Milgram & Kishino, 1994)

As shown in Figure 2, AR is located at the Reality-Virtuality (RV) continuum (Milgram & Kishino, 1994). The original Milgram RV continuum may have some ambiguity as the two ends of the continuum are not confined. Thus, a modified form of the RV continuum is presented. One end of this continuum is the real environment, where physics laws constrain the viewer in this environment and all other entities. On the other end, the virtual environment is located where the viewer is immersed in a synthetic world that may or may not exceed the physics laws' bounds, e.g., gravity or space-time continuum. Any environment between the RV continuum's extrema is defined as a Mixed Reality (MR) environment (Milgram et al., 1995). AR is a case where the viewer's vision of the environment adds the virtual contents to the real environment. In other words, in contrast to Virtual Reality (VR), AR supplements reality rather than replacing it. AR is the perfect lynchpin between the real and virtual environment and provides a natural application for a situated cognition perspective(hilken2018).

Generally, this technology's working process, like any other process, involves receiving and processing data and then providing output. The required data is received from information sources such as cameras and sensors. In the processing stage, the input information is integrated with the system's necessary information by an inbuilt algorithm and converted into the desired output. More explanation is given about the algorithms and their structure shortly. The desired information is provided to the user in the output stage according to the application context requirements. The equipment used for AR will depend on the application and its environment. Figure 3 illustrates an AR system's schematic process with the hardware usually used for input and output. This process can be described as follows.

The first step, calibration, is necessary for any system and is followed by the scene acquisition step. In this step, two data sets are given to the processor. These two sets are 1. The acceptable range of variables, names, maps, and devices models 2. Sensors output such as GPS, accelerometer, thermometer, and camera. According to the application, the number of cameras may vary. Inertial Measurement Unit (IMU) and magnetic or electromagnetic sensors are complementary and can provide higher accuracy. Also, the point cloud reconstruction is mandatory only if the depth data is processed (Oliveira et al., 2014). According to the algorithm, this data is processed, and the appropriate virtual visual content and its proper placement are prepared.

Next, the target object is identified, and if the system's goal is to present textual content, the pose identification of the target is not needed. Usually, the system is designed to present various visualizations, where pose identification and refinement are needed to align the contents with the object's pose. The algorithms used in these steps are divided into three categories:

1. Based on pattern recognition
2. Based on the position
3. SLAM (Simultaneous Localization And Mapping).

Pattern recognition-based lgorithms determine virtual image positions by identifying a barcode or sign. Newspapers, magazines, and video translation services use this algorithm. Position-based algorithms process the device's precise GPS information and process its relative position sensor. The most advanced and sophisticated algorithm, SLAM or simultaneous localization and mapping, can process AR images without the need for GPS and markers by simultaneously processing a large number of sensors such as an accelerometer and information of images received from the camera to detect the relative position of objects to each other and the device. This type of algorithm is mainly used in phones and HMDs, although it is cumbersome in processing.

Interaction with virtual contents is possible through the Interaction Handling module, which can be done by the computer, Hand Based devices, or Human-Computer Interaction (HCI) techniques.

For output displaying, there are three types of devices, according to Figure 4.

1. Mobile displays where virtual images are added via gadgets such as mobile or tablet on reality 2. Wearable devices where an HMD illustrates the virtual contents.

3. Portable displays where a projector overlays the virtual contents in the real environment.

Mobile phones and tablets are portable but generally provide mediocre images, except when the information is processed by a computer and displayed on a phone or tablet. HMDs have the right balance between the quality of virtual content and portability. Further information about HMD can be found in literature (Sahu et al., 2020).

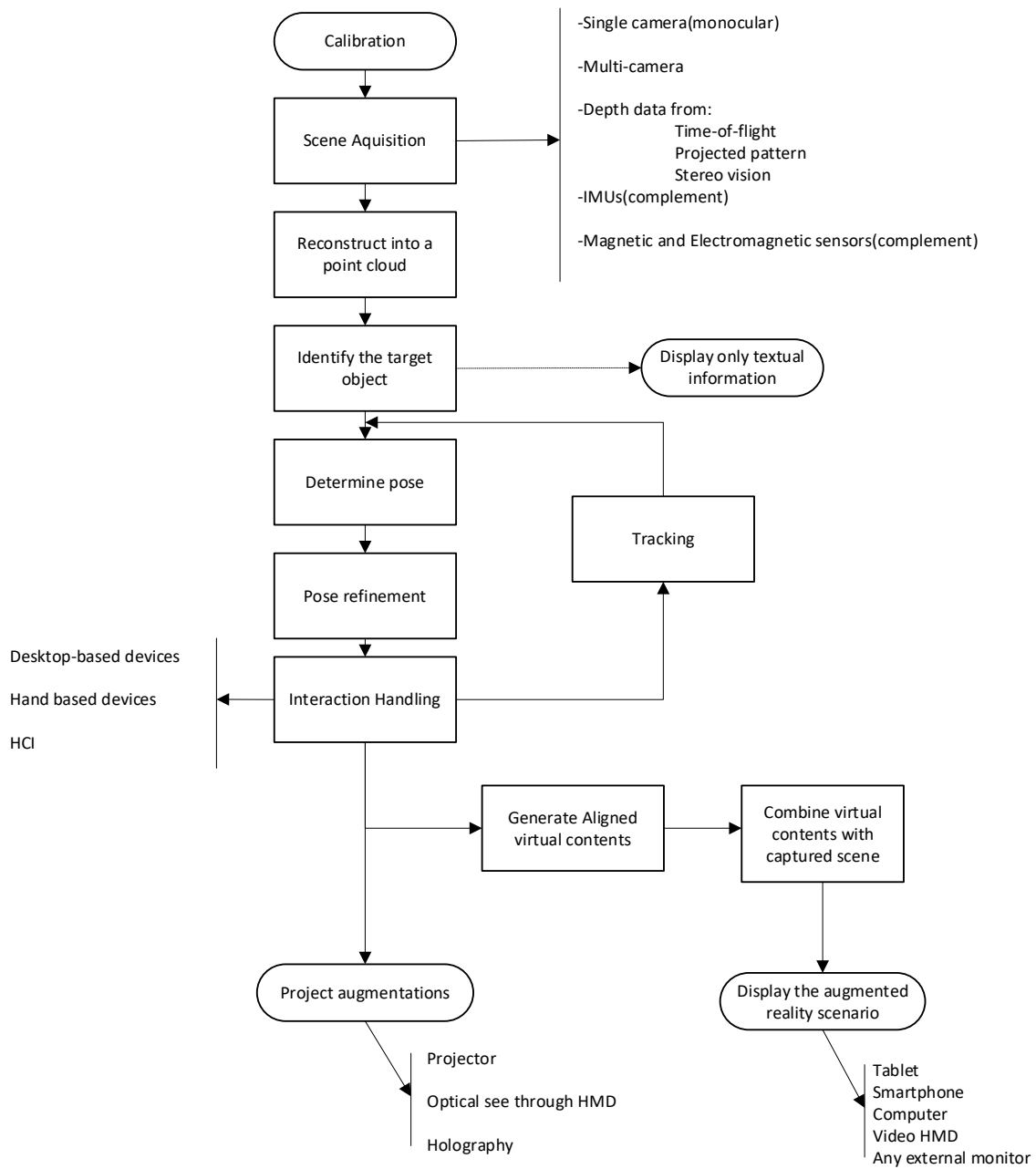

Figure 3. AR general process.

Source: Based on (Oliveira et al., 2014; Fraga-Lamas et al., 2018)

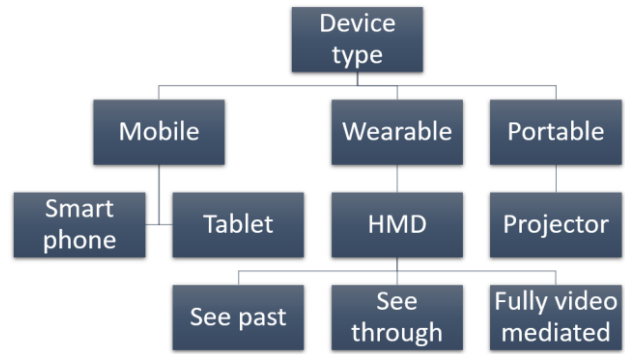

Figure 4.Various types of AR content illustration medium.

Source: reproduced from (Aleksy et al., 2014))

## 2. IAR applications for manufacturing

There are various applications reported for IAR. Figure 5 categorizes these applications into seven categories which are further divided into sub-categories. This categorization is based on the work done by this reference (Fraga-Lamas et al., 2018); however, slight modifications have been applied to include new evolved concepts in this context, mostly applied on sub-categories.

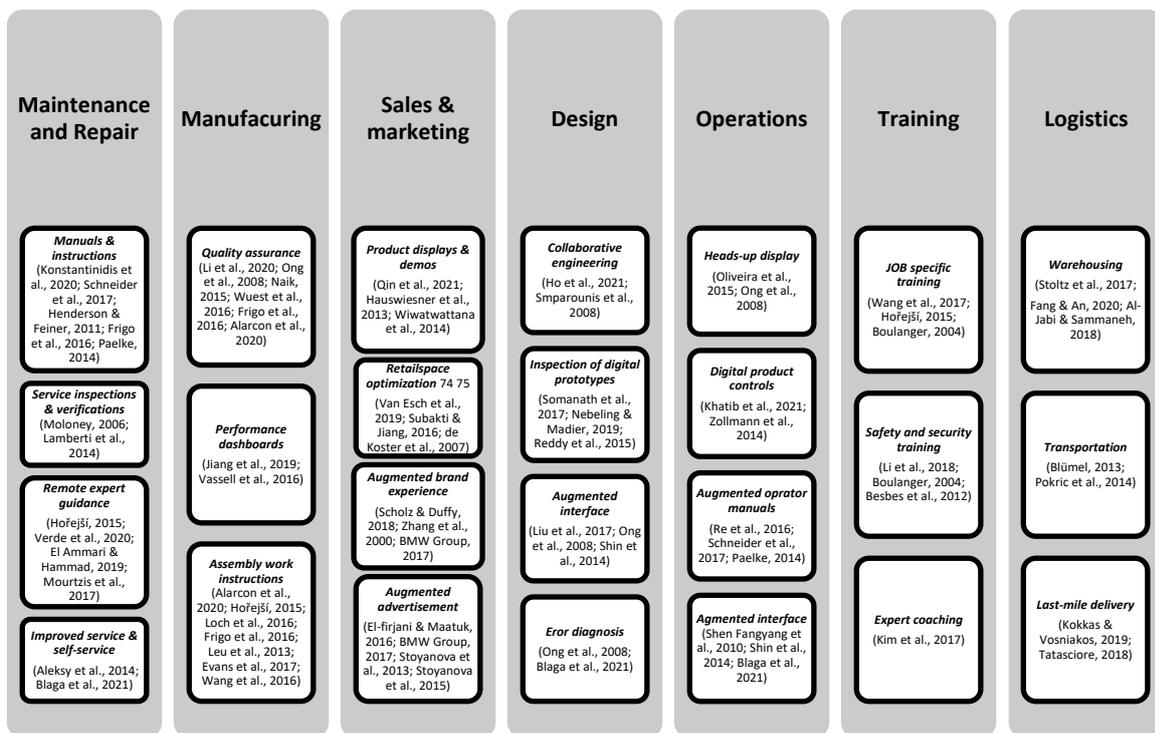

Figure 5.IAR value across I4.0.

Source: Based on (Fraga-Lamas et al., 2018))

### 2.1. Maintenance and repair

Nowadays, the maintenance process has become an essential aspect of competitiveness and productivity (Fiorentino et al., 2014). AR can offer many advantages in the repair and maintenance field, assisting workers in providing visual and auditory information (Aleksy et al., 2014; Siderska and Jadaan, 2018). When the workers are doing repair or maintenance procedures, they usually get instructions from handbooks, especially when the procedure is complicated and it is reported to be extremely time-consuming (Sanna et al., 2015; Dini & Mura, 2015; Yuan et al., 2008). When the maintenance task is complicated, like repairing a jet engine, getting instructions from a handbook is quite time-consuming compared to using an HMD to show the instructions. Working with HMDs to get the instructions from, the technician concentration would increase (Henderson & Feiner, 2009). Using HMD instead of a manual, head and eye movement reduces, leading to productivity increase (Reinhart & Patron, 2003). According to (Didier et al., 2005), AR can offer some improvements in various ways, including:

1. Transforming manuals to electronic multimedia, leading to less downtime and errors
2. Providing a novel tool for training and assisting new technicians, resulting in reduced training time.

Fiorentino et al. (2014) used AR for motorbike engine maintenance and reported that operation time was reduced by up to 79%, and the errors decreased to 92.4%.

Another maintenance-related application of IAR is "remote assistance" or "collaborative maintenance." In this way, the person calls a senior maintenance specialist or supplier of the desired part and consults with him on how to perform the repair operation, which reduces the need for experts' physical presence during the repair operation. (Wang & Liu, 2014) states that the "on-the-phone" remote assistance is inadequate for current complex maintenance tasks and mentions that AR could effectively transfer the information between expert and technician in real-time. According to the report of (Havard et al., 2015) from (Bottecchia, 2010), AR reduced the remote assistance operation time by 10% compared to phone assistance.

### 2.2. Manufacturing

For manufacturing purposes, AR can assist workers by continuously displaying the procedural information to the user and minimize their cognitive load, and they can work without altering their focus, increasing productivity and process efficiency. Also, by using AR, operators can efficiently work with new parts, and product customization can be done competently (Sahu et al., 2020).

According to (Neumann & Majoros, 1998), most manufacturing errors are related to procedural errors, while many errors are due to negligence. Errors are due to the unavailability of the right information at the right time(Alarcon et al., 2020). Thus, the object of using AR is to efficiently provide workers with the relevant information at the right time and place with appropriate quality (Reinhart & Patron, 2003). AR implementation in manufacturing processes should lead to increased productivity by reducing the completion time of the process (for both manual and cognitive tasks) and diminishing errors, resulting in reduced mental load compared to traditional media for task instruction (Alarcon et al., 2020).

Another study, (Wang et al., 2013) developed an interactive manual assembly design system, which simulates a manual assembly without the need for auxiliary computer-aided design information. (Chen et al., 2014) developed an AR system to guide the operator in the assembly of a gearbox. AR can remotely support assembly operations to facilitate customer customization requests (Mourtzis, Zogopoulos, et al., 2019).

### 2.3. Sales & marketing

With the advent of technology, novel and previously unimagined marketing choices are facilitated (Yaoyuneyong et al., 2016). The proliferation of smartphones paved the way for AR's consumer market (Ross & Harrison, 2016). The term Augmented Reality Marketing (ARM) was recently coined, describing a progressively growing marketing strategy focusing on mobile devices for AR content illustration.

ARM transforms the traditional marketing media into a hypermedia portal (Al-Modwahi et al., 2012), which allows for creative ways of experiencing product, service, and brand experience. Through ARM, the product can be put in the hands of the users, which creates an opportunity for customers to interact with the product (Al-Modwahi et al., 2012). ARM even lets consumers virtually "try on" products before purchasing them (Owyang, 2010). ARM helps marketing by 1. Increasing consumer's level of immersion, leading to enhancing communication, and 2. Enhancing sales strategy and processes (Trubow, 2011). Furthermore, (Trubow, 2011) mentions that in a study comparing traditional and AR hypermedia advertisements, a strong correlation was observed among AR technology to create immersive ads and 1. Time spent with the ad 2. Probability of purchase 3. Inclination to pay more. (Yaoyuneyong et al., 2016) anticipates that the marketers who already use ARM through smartphones will be prepared to efficiently utilize the HMDs, e.g., smart glasses, to achieve their desired prominence.

**2.4. Design**

Product design is a reciprocal-collaborative process, and several companies make much effort to increase the efficiency of this process. Observing the three-dimensional model designed during the design process can give the designer a better understanding of the model and reduce potential problems. An AR system allows designers to interact with the 3D models in real-time via their HMDs and modify them. They can also take advantage of the real environment objects for communication/interaction, e.g., stylus and whiteboard (Ong & Shen, 2009). AR enables multiple designers to simultaneously view and interact with virtual/real objects (Kim & Maher, 2008). In such an AR-based collaborative system where the designers are co-located, the collaborative design process improves (Szalavari et al., 1998). Also, when the designers are distributed in different places, AR provides more intuitive information to other collaborators' actions and operations (Sakong & Nam, 2006; Shen et al., 2008).

Additionally, AR has assisted product designers by annotating design variations(Bruno et al., 2019); in guiding operators to setup die cutters(Álvarez et al., 2019); sequence planning of assembly/disassembly(Wang, Ong, et al., 2013; Chang et al., 2020); and in planning the robot's orientation (Fang et al., 2012).

Thus, AR speeds up the design process by providing a common platform between people to observe, interact, and understand the designer's model. On the other hand, this technology eliminates prototyping for each design stage by providing an interactive virtual prototype.

**2.5. Operations**

When an operator comes across complex operations such as assemblies with numerous components and complex shapes, the operation may present challenges for the operator to accurately recognize and proceed with the process (Re et al., 2016). According to (Neumann & Majoros, 1998), the operator's interpretation and execution can require substantial mental effort. Variables such as repetitive patterns or the number/type of assembly positions can increase confusion and lead to increasing the task completion time (Richardson et al., 2004). IAR assists workers in the decision-making during the processes and operations by combining the physical experience and the information extracted in real-time from databases (Moloney, 2006). Also, IAR can provide quick access to operation manuals and drawings (Shen Fangyang et al., 2010; Henderson & Feiner, 2011; Zollmann et al., 2014). Using IAR for assisting workers during operation execution, for most cases, increased their efficiency and accuracy than a traditional, typically printed, instruction media (Re et al., 2016).

**2.6. Training**

(Abraham & Annunziata, 2017) estimated in their 2015 report that 3.5 million jobs would be created in the United States by 2025, of which 2 million would be vacant due to a lack of skills to do the job. This increases the importance of educating people to enter the world of work and educating people at the highest levels of organizations to increase their skills. IAR is believed to be instrumental in closing the skill gap, responsible for skilled workers shortage. IAR can train well-trained operators by giving step-by-step instructions for developing specific tasks (Fraga-Lamas et al., 2018). IAR is specifically useful for training machinery operators by reducing the time and effort for checking manuals (Hořejší, 2015). Thus, IAR reduces training time for new employees and lowers the skill requirements for new hires.

During the 1960s and 1970s, massive workers were hired that are currently retiring (United Nations Industrial Development Organization (UNIDO), 2013). IAR training systems can help experienced workers transfer their practical knowledge acquired by years of experience to new workers, resulting in preserving their practical knowledge (Boulanger, 2004; Besbes et al., 2012). IAR training systems can reduce the necessity of physical presence of those responsible for training by facilitating video conferences during training and task execution.

### 2.7. Logistics

The ever-growing presence of multi-brand and multi-product firms, coupled with product fragmentation, are necessary drivers for enhanced product visualization and identification (Rejeb et al., 2020). This is mainly because a visualized or digitized product, also known as a physical product digital twin, can reduce potential; production errors and non-conformance costs (Ginters & Martin-Gutierrez, 2013). Therefore, IAR, by conveniently providing precise identification and visualization of products and analysing warehouse information, can offer many logistics applications.

IAR can provide workers with virtual contents of the target object, e.g., a warehouse shelving system, and guide the worker to reach the target object in less time. For example, (Ginters & Martin-Gutierrez, 2013) developed an AR RFID system that reduces errors and damage risks by increasing the visibility of items in a warehouse. Also, firms can significantly reduce palletization errors and increase efficiency by using an AR system (Hammerschmid, 2017); such a system was proposed by (Kretschmer et al., 2018), ensuring better visualization and guidance. Also (Kretschmer et al., 2018) reports that error rates decreased significantly through AR incorporation with palletization.

Order picking is one of the most labor-intensive warehousing functions because of the time necessary to handle items and the traveling time between the locations (Tompkins, 2010). According to (Frazelle, 2001) the order picking costs can exceed fifty percent of total warehouse operating costs. IAR supports the navigation of products and reduces search time and errors like mispicks than a pick-by-voice system (Reif & Walch, 2008). IAR guidance system for order picking with an HMD is reported to outperform pick-by-light solutions. Such AR-based systems are presented by (Schwerdtfeger et al., 2009) and (Reif et al., 2010) and referred to as a pick-by-vision system and usually work with an HMD. Therefore, an AR integration for order picking can improve the process by reducing the strain on the users, minimizing errors, and increasing productivity (Rejeb et al., 2020).

For the transportation industry, optimized load planning and vehicle utilization rely heavily on digital data and planning software (DHL, 2014). In this regard, AR can help by reducing the required time for package identification, optimized truck sequencing and loading, and routing strategies (Rejeb et al., 2020). The system proposed by (Ginters & Martin-Gutierrez, 2013) displays container information and other packed objects by combining RFID and AR technologies. Using AR, several problems associated with last-mile delivery, such as the difficulty of locating specific products within a delivery vehicle, could be addressed by visualizing the required information, leading to a more efficient logistics process.

### 3. IAR challenges and research opportunities

Despite its numerous advantages, IAR faces many challenges for being fully implemented in the industry. (Egger & Masood, 2019) and (de Souza Cardoso et al., 2020) conducted a systematic literature review (SLR) through IAR-related papers. All challenges presented in Figure 6, except the hardware and user's health and acceptance, are directly or indirectly related to the AR software. This is in conformance with (Egger & Masood, 2019) claim that among the various challenges that IAR faces to be fully implemented into industries, the main challenges are related to its technology, especially its software (Figure 7). JABIL conducted a survey among technology and business stakeholders in product companies concerning AR and VR challenges. In their survey, 99% stated that AR faces technology challenges. In comparison, 61% think that AR needs innovation, meaning that there is still much work to be done on technology innovation instead of investment to become mainstream in the near future (JABIL, 2018).

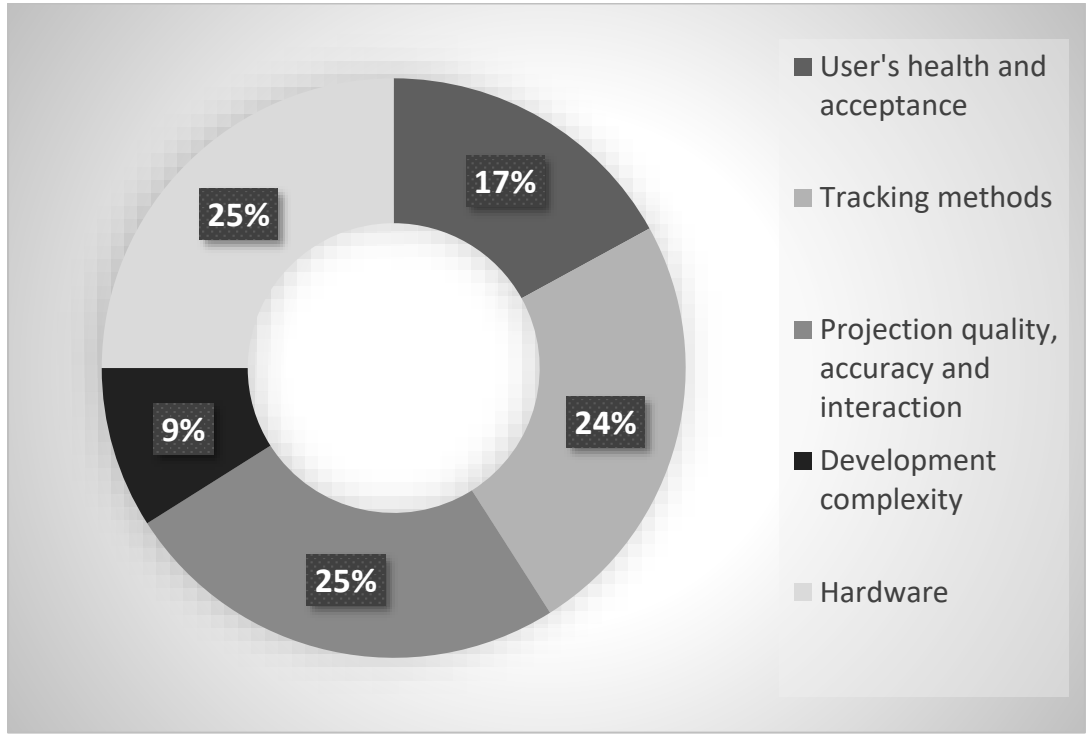

Figure 6. IAR main challenges.

Source: Reproduced from (de Souza Cardoso et al., 2020)).

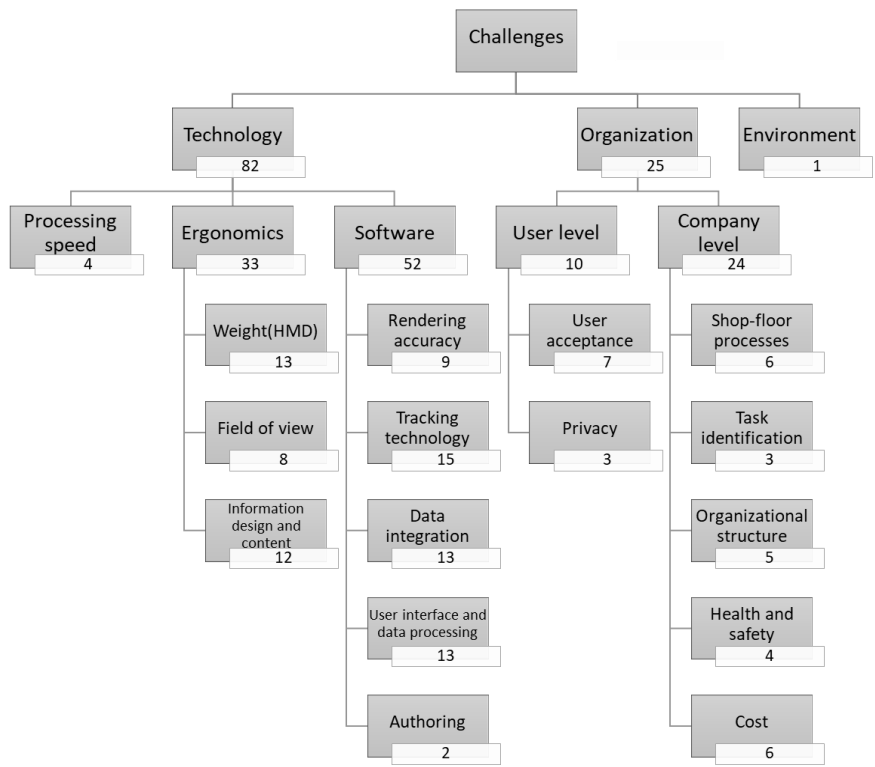

Figure 7. IAR main challenges.

Source: Reproduced from (Egger & Masood, 2019)).

### 3.1. Technology Challenges
#### 3.1.1. Processing speed

As reported by (Porcelli et al., 2013) and (Stoltz et al., 2017), the processing power of AR devices, especially HMDs, is not adequate, causing poor user experience and performance. High processing power is needed for AR devices to perform real-time scenarios with high-quality digital content and tracking algorithms. Although Moore laws suggest that approximately every two years, the density of transistors on integrated circuits doubles (Moore, 2006), meaning that the processing power of the AR devices increases in time, but also the amount of the processing power demands increases as AR further gets implemented in industrial scenarios. As suggested by (Huang et al., 2012), cloud computing seems an appropriate solution as it offers better resource and content utilization. However, it needs high-quality network service and security to ensure smooth content delivery (Aziz et al., 2012). Such a cloud-based AR system is presented by (Fernández-Caramés et al., 2018). (Seam et al., 2017) states that cloud-based AR systems need high network capability to transform data from the cloud to the visualization device. As suggested by Figure 8, 5G internet network offers adequately low-latency and high-bandwidth mobile connection for AR to operate appropriately.

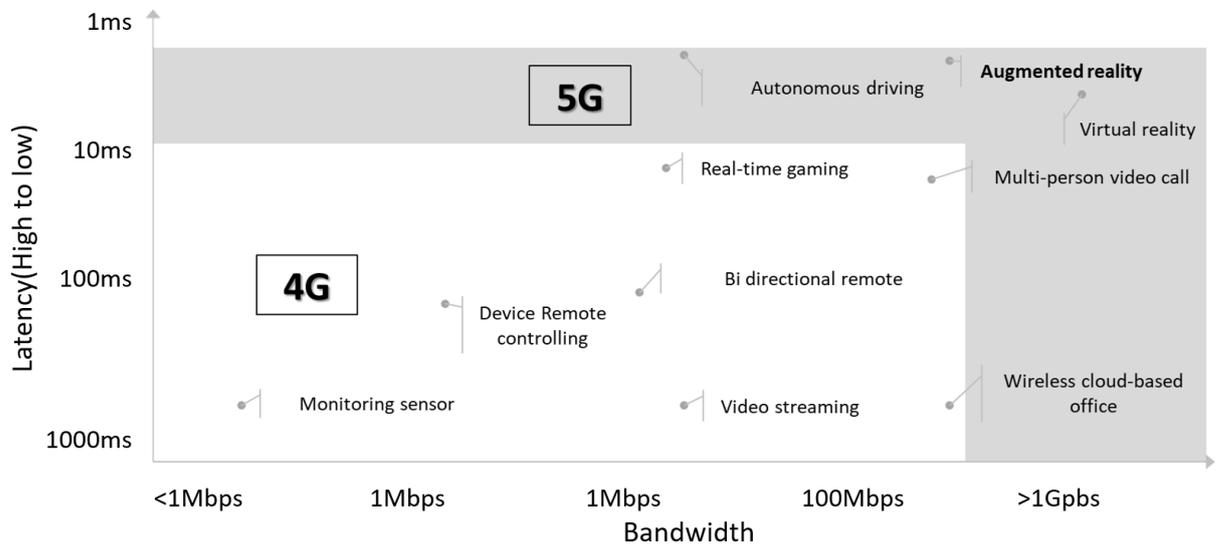

Figure 8. Projected network demands for emerging technologies.

Source: Based on (Seam et al., 2017))

#### 3.1.2. Ergonomics

The user acceptance of a product is connected with its ergonomic design. Not only its hardware, namely, weight and FOV was reported as a problem (Palmarini et al., 2018; Holm et al., 2017; Schlagowski et al., 2017; Makris et al., 2016; Maly et al., 2016; Mura et al., 2016), but also its software ergonomics like light sensitivity of AR algorithms and camera angles were criticized (Akçayır & Akçayır, 2017; Chang et al., 2015). Long-term usage of AR devices, especially for HMDs) was reported to cause distorted depth perception, diplopic vision, visual discomfort, and fatigue (Murauer et al., 2018; Hua & Javidi, 2015). Although using AR reduces eye and neck movement during operations (Henderson & Feiner, 2011), operators can get distracted or disoriented by the virtual contents (Funk et al., 2017; Hou et al., 2013).

#### 3.1.3. Software

Marker-based algorithms are not ideal for many industrial applications as they are often light-sensitive, and the marker can be obstructed by components or tool shadows (Fang et al., 2014). Their size and potion are potential barriers during task execution (Liu & Seipel, 2016). Markerless methods are yet immature and require a long setup time since they have difficulties mapping the environment (Soete et al., 2015) and virtual objects instabilities (Hou et al., 2014). Despite the recent progress in deep learning, the artificial intelligence that supports AR applications is still in its infancy (Interrante et al., 2018). Object tracking algorithms are often reported as a problem (Carmigniani et al., 2010). AR content quality also needs improvements (Palmarini et al., 2018).

As this technology matures and gets further implemented in the industry, its data structures, modeling, and interfaces need to follow specific standards like UML or OPC UA (Flatt et al., 2015; Havard et al., 2015) to facilitate their integration. It is not yet clear which standard from the numerous standards (Trappey et al., 2017) will dominate each application. Further information on the standardization of AR can be found in (Garzón et al., 2017).

### 3.2. Organization challenges
#### 3.2.1. User-level

Some user-related challenges were discussed in the ergonomics section, yet user acceptance is also related to privacy and protection (Roesner et al., 2014; Stoltz et al., 2017; Syberfeldt et al., 2016; Hou et al., 2015). Training evaluation and task/error tacking (Wolfartsberger et al., 2017) require monitoring the AR users and their gestures, voice, and actions monitored by the AR system (De Pace et al., 2018), increasing superior surveillance.

#### 3.2.2. Company-level

Some possible challenges or incompatibilities were reported when testing using AR in the industry (Funk et al., 2017; Espíndola et al., 2013; Espíndola, Pereira, et al., 2013; Garza et al., 2013; Gavish et al., 2013; Porcelli et al., 2013; Real & Marcelino, 2011). Some field studies reported a negative influence of AR on performance due to the comfortability of HMDs, Software challenges, privacy issues, and high cognitive load(Stoltz et al., 2017; Yuviler-Gavish et al., 2011). Also, it is indicated that the task's complexity level is correlated to AR utilization(Blattgerste et al., 2017; Syberfeldt et al., 2016; Yuviler-Gavish et al., 2011). Health and safety concerns (Murauer et al., 2018; Makris et al., 2016) due to distraction and cost or profitability implications was also reported (Stoltz et al. 2017; Hou et al., 2015; Espíndola, Pereira, et al., 2013; Gavish et al., 2013; Porcelli et al., 2013; Bondrea & Petruse, 2012). Like any other network, AR systems are susceptible to potential cyber attackers. Thus, companies might face some cybersecurity challenges to protect their individual intellectual property (IP) and the integrity of machinery and processes (Langfinger et al., 2017).

### 3.3. Environment challenges

According to (Egger & Masood, 2019), the environment of a company implementing IAR might play a role in how the technology is used. These challenges might include third-party information integration, the industry-wide standardization of AR solutions, the employment regulatory environment, or the external support necessity as mentioned by (Stoltz et al. 2017). It is also needed to analyse the emerging behvaiour of using the AR in a system to understand better its environmental impact at a higher level.

### 3.4. Research opportunities

Many researches have been done to address the aforementioned challenges. An emerging technology known as metalens has recently become popular. This technology basically operates on the same principle that conventional diffractive lenses do. Although they seem similar, the metalenses can have unique properties like high numerical aperture capability and tunability, meaning a wider FOV and zooming ability for the HMD (Engelberg & Levy, 2020).

Researchers are working on haptic feedback for virtual objects by producing an illusion of touching the virtual objects, resulting in the user's sense of object presence. Object presence does not necessarily mean the real environment's presence; instead, it means the user subjectively experiences the virtual object's feeling to exist in the real environment, even when the object does not (Stevens & Jerrams-Smith, 2000). Also, cognitive activity is made possible through interacting with an object with visual, auditory, and haptic sensory simultaneously (Gibson, 2015). The research on AR's sensory stimuli is more promising compared to those of VR, mostly where the reality augmentations are done by an HMD (Kang & Lee, 2018). Touch Simulation of an object has two stages: 1. Convert touch data to signal(digitization), which can be done 2. Simulate touch on the skin.

The touch of objects is modeled in tactile and touchless modes. The tactile mode is based on changing the material's electrical resistance due to the distance or proximity of two conductive plates to each other. As the two materials approach each other, the sensor's signal (the blue part of the graph) is amplified. The touchless mode is a state in which a person works with a sensor through magnets on their fingertips. In this state, as the magnetic material approaches the sensor's surface, the amount of signal is attenuated, indicating that the magnetic material is approaching. For both modes, when these sensors form an array and calibrated, they can measure the position and intensity of touch with high accuracy (Cañón Bermúdez et al., 2018; Ge et al., 2019; Yu et al., 2019).

For simulating skin touch, pneumatic circuits are used that can apply a force with high accuracy in both fixed and vibrating states (Sonar et al., 2020; Frediani & Carpi, 2020). In this way, the amount of tactile force, softness or stiffness of the body, and simulation of vibrating objects can be achieved from this method.

Using the N * N arrays of the introduced sensors and actuators in the evolved state, they can be positioned to perceive the touch information in three dimensions and simulate the touch on numerous skin points. Thus the third sense of the five senses is added to AR technology.

## 4. Conclusion

AR already proved its positive role in many industries such as manufacturing and logistics, continuing to expand in the future. Also, AR paves the way for other I4.0 technologies such as blockchain, IoT, and cloud computing to better integrate through firms and increase other technologies' efficiency (Blümel, 2013). However, further IAR implementation may face the challenges presented in this study, while some of the challenges may require more developments on other pioneering technologies like advanced optics and cloud computation. With the promising trend of IAR, decision-makers in companies should be informed about the influence of AR on their enterprises and develop initial plans or frameworks for AR implementation, if possible.

Times New Roman 11, spacing: at least 13 pt.
Times New Roman 11, spacing: at least 13 pt.